\documentclass[aps,prl,reprint,groupedaddress]{revtex4-1} 
\usepackage[colorlinks=true,urlcolor=blue]{hyperref}
\usepackage{amssymb,amsmath,amsfonts}
\usepackage{epsfig}
\usepackage{graphicx}
\usepackage{epstopdf}

\def\clock{{\count0=\time
           \divide\count0 60
           \ifnum\count0<10 0\fi\the\count0
           \multiply\count0 -60 \advance\count0 \time
           :\ifnum\count0<10 0\fi \the\count0
         }}
\newcommand{\timestamp}{{\small\vbox{\hbox{\tt\jobname.tex}
\hbox{\the\day/\the\month/\the\year, \clock}}}}

\newcommand{\be}{\begin{eqnarray}}
\newcommand{\ee}{\end{eqnarray}}
\newcommand{\beq}{\begin{eqnarray}}
\newcommand{\eeq}{\end{eqnarray}}

\newcommand{\beqa}{\begin{eqnarray}}
\newcommand{\eeqa}{\end{eqnarray}}

\let\oldsqrt\sqrt
\def\sqrt{\mathpalette\DHLhksqrt}
\def\DHLhksqrt#1#2{%
\setbox0=\hbox{$#1\oldsqrt{#2\,}$}\dimen0=\ht0
\advance\dimen0-0.2\ht0
\setbox2=\hbox{\vrule height\ht0 depth -\dimen0}%
{\box0\lower0.4pt\box2}}

\allowdisplaybreaks

\begin{document}

\rightline{\vbox{CCTP-2014-17, CCQCN-2014-40  \phantom{ghost} }}

\title{Schr\"odinger Invariance from Lifshitz Isometries in Holography and Field Theory}

\author{Jelle Hartong$^1$, Elias Kiritsis$^{2,3}$,  and Niels A. Obers$^1$}
\email[]{hartong@nbi.dk, kiritsis@physics.uoc.gr, obers@nbi.dk}
\affiliation{$^1$The Niels Bohr Institute, University of Copenhagen,
Blegdamsvej 17, 2100 Copenhagen \O, Denmark, }
\affiliation{$^2$ Crete Center for Theoretical Physics, Department of Physics, University of Crete,
 71003 Heraklion, Greece,}
\affiliation{$^3$ APC, Universit\'e Paris 7, CNRS/IN2P3, CEA/IRFU, Obs. de Paris, Sorbonne Paris Cit\'e,
B\^atiment Condorcet, F-75205, Paris Cedex 13, France, (UMR du CNRS 7164).}
.



\begin{abstract}

We study non-relativistic field theory coupled to a torsional Newton--Cartan geometry both directly as well as holographically. The latter involves gravity on asymptotically locally Lifshitz space-times. We define an energy-momentum tensor and a mass current and study the relation between conserved currents and conformal Killing vectors for flat Newton--Cartan backgrounds. It is shown that flat NC space-time realizes two copies of the Lifshitz algebra that together form a Schr\"odinger algebra (without the central element). We show why the Schr\"odinger scalar model has both copies as symmetries and the Lifshitz scalar model only one. Finally we discuss the holographic dual of this phenomenon by showing that the bulk Lifshitz space-time realizes the same two copies of the Lifshitz algebra.

\end{abstract}

\pacs{}

\maketitle


\noindent\textbf{Introduction.} One of the corner stones of theoretical physics is the relation between space-time symmetries and conservation laws. In relativistic field theories it is well-known how to obtain a conserved current for each isometry of the background Lorentzian geometry. Much less is known about the precise manner in which to implement such ideas in the realm of non-relativistic field theories. Such theories are naturally formulated on a torsional Newton--Cartan (TNC) background which is a generalization of Newton-Cartan (NC) geometry that allows for torsion. We will show that for field theories on a TNC background the interplay between conserved currents and space-time isometries is markedly different from the relativistic case involving a new mechanism. 

Field theory on TNC backgrounds has recently appeared in studies of systems with strongly correlated  electrons, such as the quantum Hall effect \cite{Gromov:2014vla,Geracie:2014nka,Brauner:2014jaa,Geracie:2014zha} following the earlier work \cite{Son:2013rqa} that suggested to use NC geometry in this context. It was recently found that the boundary geometry of asymptotically locally Lifshitz space-times is described by TNC geometry \cite{Christensen:2013lma,Christensen:2013rfa,Hartong:2014oma,Bergshoeff:2014uea}. 

We first discuss the relation between conserved currents and isometries in the context of perhaps the simplest of all non-relativistic field theories namely the one giving rise to the Schr\"odinger equation. To the best of our knowledge the new perspective presented here on the Schr\"odinger symmetry of the Schr\"odinger equation, which relies on formulating it as a field theory on a TNC background, has not appeared elsewhere. Then we show that the same mechanism is at work in holographic dualities between field theories on TNC backgrounds and their gravitational duals defined on asymptotically locally Lifshitz space-times. We will study the space-time symmetries of a bulk Lifshitz space-time and show that certain bulk diffeomorphisms relate different Lifshitz subalgebras of the same Schr\"odinger algebra. This provides another perspective on the results of \cite{Hartong:2014oma,Bergshoeff:2014uea} where it is shown that the sources and vevs of asymptotically locally Lifshitz spacetimes transform under the Schr\"odinger algebra.

\noindent\textbf{Torsional Newton--Cartan Geometry.} We start our discussion with a succint summary of torsional Newton--Cartan (TNC) geometry as formulated in \cite{Bergshoeff:2014uea}. The geometry can be described in terms of the following fields: the vielbeins $\tau_\mu$ and $e^a_\mu$ ($\mu$ is a $(d+1)$-dimensional space-time index and $a=1,\ldots,d$ a spatial tangent space index), the vector field $M_\mu$ and the St\"uckelberg scalar $\chi$. The latter appears only via the St\"uckelberg decomposition $M_\mu=\tilde m_\mu-\partial_\mu\chi$ which defines $\tilde m_\mu$ given $M_\mu$ and $\chi$. These fields transform under diffeomorphisms ($\xi^\mu$), local dilatations $D$ ($\Lambda_D$), Galilean boosts $G_a$ ($\lambda^a$), rotations $J_{ab}$ ($\lambda^{ab}$) and gauge transformations $N$ ($\sigma$) as 
\begin{equation}\label{eq:gaugetrafos}
\begin{array}{rcl}
\delta \tau_\mu & = & \mathcal{L}_\xi\tau_\mu+z \Lambda_D \tau_\mu \,,  \\
\delta e_\mu^a & = & \mathcal{L}_\xi e^a_\mu+\lambda^a \tau_\mu + \lambda^{a}{}_b e_{\mu}^{b} + \Lambda_D e_\mu^a \,, \\
\delta M_\mu & = & \mathcal{L}_\xi M_\mu+ e_\mu^a\lambda_a+(2-z)\Lambda_D M_\mu\,,\\
\delta\chi & = & \mathcal{L}_\xi\chi+\sigma+(2-z)\Lambda_D\chi\,,
\end{array}
\end{equation}
where $z>1$. We define inverse vielbeins $v^\mu$ and $e^\mu_a$ via
\begin{equation}\label{eq:vielbeins}
\begin{array}{rclrcl}
v^\mu \tau_\mu & = & -1\,, & \hskip .2cm  v^\mu e_\mu^a & = & 0 \,, \\
 \tau_\mu e^\mu_a & = & 0\,, & \hskip .2cm e_\mu^a e^\mu_b & = & \delta^a_b\,.
\end{array}
\end{equation}
We have the completeness relation $e^\mu_a e^a_\nu = \delta^\mu_\nu + v^\mu \tau_\nu$.

The first step in setting up the TNC geometry is the construction of invariants, i.e. tensors with a specific dilatation weight that are invariant under $G$, $J$, $N$ transformations. These invariants are given by
\begin{equation}\label{eq:tildePhi}
\begin{array}{rcl}
\hat v^\mu & = & v^\mu - h^{\mu \nu} M_\nu \,, \\ 
\bar h_{\mu\nu} & = & h_{\mu\nu}-\tau_\mu M_\nu-\tau_\nu M_\mu\,,\\
\tilde\Phi & = & -v^\mu M_\mu+\frac{1}{2}h^{\mu\nu}M_\mu M_\nu\,,
\end{array}
\end{equation}
together with the degenerate metric invariants $\tau_\mu$ and $h^{\mu\nu}$. The scalar $\tilde\Phi$ is closely related to the Newtonian potential \cite{Bergshoeff:2014uea}. It will also sometimes be useful to use the $G$ and $N$ invariant vielbein $\hat e_\mu^a$ defined as $\hat e_\mu^a = e_\mu^a-\tau_\mu M^a$. The objects $\hat e_\mu^a$, $\hat v^\mu$, $\tau_\mu$, $e_\mu^a$ form an orthonormal set.

The $G$, $J$, $N$ invariant affine connection that is metric compatible in the sense that
\begin{equation}\label{eq:TNC}
\nabla_\mu\tau_\nu = 0\,,\hskip 1cm\nabla_\mu h^{\nu\rho} = 0\,,
\end{equation}
is given by
\begin{equation}\label{eq:GammaTNC}
\Gamma^{\rho}_{\mu\nu} = -\hat v^\rho\partial_\mu\tau_\nu+\frac{1}{2}h^{\rho\sigma}\left(\partial_\mu\bar h_{\nu\sigma}+\partial_\nu \bar h_{\mu\sigma}-\partial_\sigma\bar h_{\mu\nu}\right)\,.
\end{equation}

\noindent\textbf{Flat NC Space-Time.} An important role will be played by the notion of flat Newton--Cartan space-time which we will define next (see also \cite{Andringa:2013zja,Bergshoeff:2014uea}). There exists a coordinate system, $x^\mu=(t,x^i)$, referred to as a global inertial coordinate system, for which we have
\begin{equation}\label{eq:flatNC}
\begin{array}{rclrclrcl}
\tau_\mu & = & \delta^t_\mu\,,&\hskip .2cm h^{ij} & = & \delta^{ij}\,,\\
v^\mu & = & -\delta^\mu_t\,,&\hskip .2cm M_\mu & = & \partial_\mu M\,.&&&
\end{array}
\end{equation}
Here $t$ is absolute time, $x^i$ are Cartesian spatial coordinates, the choice for $v^\mu$ fixes the freedom to perform local Galilean boosts and $M_\mu = \partial_\mu M$ ensures that the connection $\Gamma^\rho_{\mu\nu}$ is everywhere zero, so that we restrict to inertial observers. We furthermore impose that $M=\text{cst}$ up to those local transformation \eqref{eq:gaugetrafos} that preserve the choices \eqref{eq:flatNC}. This implies that we also have
\begin{equation}\label{eq:noNP}
\tilde\Phi=\partial_t M+\frac{1}{2}\delta^{ij}\partial_i M\partial_j M=0\,,
\end{equation}
as well as two other conditions that tell us that $M$ can be at most trace quadratic in $x^i$, i.e. $M=ax^i x^i+b^ix^i+c$ where the coefficients are functions of $t$ \cite{Hartong:2015wxa}. Hence, flat NC space-time in global inertial coordinates comes together with a trace quadratic function $M$ obeying \eqref{eq:noNP}.

\noindent\textbf{Conformal Killing Vectors of Flat NC Space-Time.} To contrast flat NC space-time with Minkowski space-time we look at its conformal Killing vectors. The TNC conformal Killing equations are \cite{Hartong:2015wxa,Hartong:2014} 
\begin{equation}
\begin{array}{rclrcl}
\hskip -.3cm \mathcal{L}_{K}\tau_\mu & = & -z\Omega\tau_\mu\,,\hskip .3cm &\mathcal{L}_{K}\hat v^\mu & = & z\Omega\hat  v^\mu\,,\\
\hskip -.3cm \mathcal{L}_{K}\bar h_{\mu\nu} & = & -2\Omega\bar h_{\mu\nu}\,,\hskip .3cm &\mathcal{L}_{K} h^{\mu\nu} & = & 2\Omega h^{\mu\nu}\,,\\
\hskip -.3cm K^\mu\partial_\mu\tilde\Phi & = & 2(z-1)\Omega\tilde\Phi\,,& &  & 
\end{array}
\end{equation}
where $\Omega$ is any function. Substituting into these equations the flat NC conditions \eqref{eq:flatNC} it can be shown that for  $(z-2)\partial_t\Omega=0$ the conformal Killing vectors become
\begin{equation}\label{eq:K}
\begin{array}{rcl}
K^t & = & a-z\lambda t-\delta_{z,2}\alpha t^2\,,\\
K^i & = & a^i+v^i t+\lambda^i{}_j x^j-\lambda x^i-\delta_{z,2}\alpha tx^i\,,
\end{array}
\end{equation}
provided we can solve
\begin{equation}\label{eq:KE}
\mathcal{L}_{K}M = v^ix^i-\frac{1}{2}\delta_{z,2}\alpha x^ix^i+(z-2)\lambda M+C\,.
\end{equation}
The case with $(z-2)\partial_t\Omega\neq 0$ will be discussed later.

We first consider the trivial solution of \eqref{eq:noNP} given by
\begin{equation}\label{eq:choice1M}
M=\text{cst}\,.
\end{equation}
In this case the Killing vectors form the Lifshitz Lie algebra $H$, $P_i$, $J_{ij}$, $D$ that is a subgroup of the Schr\"odinger algebra for general $z$  given by
\begin{equation}\label{eq:HPJD}
\begin{array}{rclrcl}
\hskip -.2cm H & = & \partial_t\,,& \hskip .2cm  P_i & = & \partial_i\,,\\
\hskip -.2cm J_{ij} & = & x_i\partial_j-x_j\partial_i\,,& \hskip .2cm  D & = & zt\partial_t+x^i\partial_i\,.
\end{array}
\end{equation}

If on the other hand we take the solution of \eqref{eq:noNP},
\begin{equation}\label{eq:choice2M}
M = \frac{x^ix^i}{2t}\,,
\end{equation}
we get the Killing vectors
\begin{equation}\label{eq:KGJD}
\begin{array}{rclrcl}
K & = & t^z\partial_t+t^{z-1}x^i\partial_i\,,&\hskip .2cm G_i & = & t\partial_i\,,\\
J_{ij} & = & x_i\partial_j-x_j\partial_i\,,&\hskip .2cm D & = & zt\partial_t+x^i\partial_i\,.
\end{array}
\end{equation}
Actually from equations \eqref{eq:K} and \eqref{eq:KE} we only find the solution with $\alpha\neq 0$ for $z=2$ which is due to the condition $(z-2)\partial_t\Omega=0$ used in obtaining \eqref{eq:K} and \eqref{eq:KE}. When we take $(z-2)\partial_t\Omega\neq 0$ it can be shown that the conformal Killing equations allow for $K$ to be a conformal Killing vector even when $z\neq 2$. The nonzero commutators are
\begin{equation}\label{eq:Lifshitzalgebra2}
[D,K]=z(z-1)K\,,\qquad[D,G_i]=(z-1)G_i\,,
\end{equation}
where we left out the ones involving $J_{ij}$. The algebra of $D$, $K$, $G_i$ and $J_{ij}$ is isomorphic to the Lifshitz algebra.

The Schr\"odinger algebra (without the central element $N$) is obtained if we can combine all Killing vectors obtained for these two different choices of $M$. If we do this for $z\neq 2$ we loose the $K$ generator as it does not form an algebra with $H$ and $P_i$ in agreement with the well-known fact that one cannot add a special conformal generator to the Schr\"odinger algebra for $z\neq 2$. The central element $N$ corresponds to shifting $M$ by an arbitrary constant. This generator is realized on fields but does not come about as a space-time symmetry as we will see shortly.

We stress that there is no choice for $M$ that admits the entire Schr\"odinger algebra as its conformal Killing vectors. Lifshitz is the largest possible algebra that the conformal Killing vectors can span for any $M$.

\noindent\textbf{Field Theory on TNC Backgrounds.} We will next study a scalar field theory on a TNC background and define objects like the energy-momentum tensor as well as examine the role of $M$ for the case of a flat NC background. When coupling a field theory to the background fields $v^\mu$, $e^\mu_a$, $\tilde m_\mu=\tilde m_0\tau_\mu+\tilde m_a e^a_\mu$ and $\chi$ we define the following objects (vevs) when varying these fields (sources)
\begin{eqnarray}
\delta_{\text{backgrd}} S & = & \int d^{d+1}x e\left[-S^0_\mu\delta v^\mu+S^a_\mu\delta e^\mu_a+T^0\delta\tilde m_0\right.\nonumber\\
&&\left.+T^a\delta\tilde m_a+\langle O_\chi\rangle\delta\chi\right]\,,
\end{eqnarray}
where $e$ is the determinant of the matrix $(\tau_\mu\,,e_\mu^a)$. Just like for the TNC geometry it is useful to find invariants, i.e. $G$, $J$, $N$ invariant quantities that transform as tensors with a specific dilatation weight (up to possible terms involving derivatives of $\Lambda_D$). In \cite{Hartong:2014oma,Hartong:2015wxa} we show that these invariants are the energy-momentum tensor $T^\mu{}_\nu$ and mass current $T^\mu$ defined via
\begin{equation}
\begin{array}{rcl}
T^\mu{}_\nu & = & -\left(S^0_\nu+T^0\partial_\nu\chi\right)v^\mu+\left(S^a_\nu+T^a\partial_\nu\chi\right)e^\mu_a\,,\\
T^\mu & = & -T^0 v^\mu+T^a e_a^\mu\,.
\end{array}
\end{equation}
We note that in TNC geometry there is no metric that can be used to raise and lower space-time indices, so that the energy-momentum tensor is a mixed $(1,1)$ tensor. The vielbein components of $ T^\mu{}_\nu$ are obtained by contraction with $e_\mu{}^a$, $v^\mu$, $\tau_\mu$, $e_\mu{}^a$ and provide the energy density, energy flux, momentum density and stress, while $T^0$ is the mass density and $T^a$ the mass flux.

Consider the following simple example of a field theory defined on a $(d+1)$-dimensional fixed TNC background whose equations of motion give rise to the Schr\"odinger equation with potential $V$,
\begin{eqnarray}
S & = & \int d^{d+1}x e\left(-i\phi^*\hat v^\mu\partial_\mu\phi+i\phi\hat v^\mu\partial_\mu\phi^*-h^{\mu\nu}\partial_\mu\phi\partial_\nu\phi^*\right.\nonumber\\
&&\left.-2\phi\phi^*\tilde\Phi-V(\phi,\phi^*)\right)\,.\label{eq:Schaction}
\end{eqnarray}
We find the following energy-momentum tensor and mass current
\begin{equation}\label{eq:EMtensor+masscurrent}
\begin{array}{rcl}
T^\mu{}_\nu & = & -e^{-1}\mathcal{L}\delta^\mu_\nu-2h^{\mu\rho}\partial_{(\nu}\phi\partial_{\rho)}\phi^*\\
&&-\hat v^\mu\left(i\phi^*\partial_\nu\phi-i\phi\partial_\nu\phi^*\right)\,,\\
T^\mu & = & 2\phi\phi^*\hat v^\mu+h^{\mu\nu}\left(i\phi^*\partial_\nu\phi-i\phi\partial_\nu\phi^*\right)\,.
\end{array}
\end{equation}
One can check that the Ward identities 
\begin{equation}
\begin{array}{rcl}
0 & = & -T^\mu\hat e_\mu^a+T^\mu{}_\nu\tau_\mu e^{\nu a}\,,\\
0 & = & T^\mu{}_\nu\hat e_{\mu}^a e^{\nu b}-T^\mu{}_\nu\hat e_\mu^b e^{\nu a}\,,\\
\langle O_\chi\rangle & = & e^{-1}\partial_\mu\left(e T^\mu\right)\,,
\end{array}
\end{equation}
for local $G$, $J$, $N$ invariance are satisfied off-shell. When $V=V(\phi\phi^\star)$ the theory has the local symmetry
\begin{equation}
\delta M_\mu = \partial_\mu\alpha\,,\hskip 1cm \delta\phi = -i\alpha\phi\,,
\end{equation}
leading to the on-shell Ward identity $\partial_\mu\left(e T^\mu\right)=0$. The diffeomorphism Ward identity is given by
\begin{equation}
e^{-1}\partial_\mu\left(eT^\mu{}_\nu\right)+T^\rho{}_\mu\left(\hat v^\mu\partial_\nu\tau_\rho-e_a^\mu\partial_\nu\hat e_\rho^a\right)+T^0\partial_\nu\tilde\Phi=0\,.
\end{equation}
If we assume that $\phi$ has dilatation weight $d/2$ and the potential has dilatation weight $d+2$, which e.g. is the case for $V=(\phi\phi^*)^{(d+2)/d}$, it can be shown that for NC backgrounds, i.e. those for which $d\tau=0$, we have the following trace Ward identity \cite{Hartong:2015wxa}
\begin{equation}
-2\tau_\mu\hat v^\nu T^\mu{}_\nu+\hat e_\mu^ae_a^\nu T^\mu{}_\nu+2T^0\tilde\Phi=e^{-1}\partial_\mu\left(eV^\mu\right)\,,
\end{equation}
where $V^\mu$ is the virial current given by
\begin{equation}\label{eq:virialcurrent}
V^\mu=\frac{d}{2}h^{\mu\nu}\left(\phi^*\partial_\nu\phi+\phi\partial_\nu\phi^*\right)\,.
\end{equation}

Substituting the choices \eqref{eq:flatNC} for a flat NC background into the action \eqref{eq:Schaction} we get
\begin{eqnarray}
S & = &\int d^{d+1}x \left(i\phi^*\left(\partial_t\phi+i\phi\partial_t M\right)-i\phi\left(\partial_t\phi^*-i\phi^*\partial_t M\right)\right.\nonumber\\
 &&\hskip -.7cm\left.-\delta^{ij}\left(\partial_i\phi+i\phi\partial_i M\right)\left(\partial_j\phi^*-i\phi^*\partial_j M\right)-V(\phi,\phi^*)\right)\,.\label{eq:SchactionflatNC}
\end{eqnarray}
If we make the field redefinition $\phi=e^{-iM}\psi$ the action \eqref{eq:SchactionflatNC}, for potentials of the form $V=V(\phi\phi^*)$, becomes
\begin{equation}\label{eq:Schaction2}
S = \int d^{d+1}x\left(i\psi^*\partial_t\psi-i\psi\partial_t\psi^*-\delta^{ij}\partial_i\psi\partial_j\psi^*-V\right)\,,
\end{equation}
so that we can remove $M$ by a field redefinition. Variation of  \eqref{eq:SchactionflatNC} with respect to $M$ gives $\partial_\mu\left(e T^\mu\right)=0$ which when written in terms of the wavefunction $\psi$ is the conservation of probability equation of quantum mechanics.

The action \eqref{eq:SchactionflatNC} is scale invariant provided $V$ has dilatation weight $d+2$ and $\phi$ has dilatation weight $d/2$ and $M$ does not transform. Further it is invariant under special conformal transformations given by
\begin{equation}
\begin{array}{rclrcl}
t & = & \frac{t'}{1-ct'}\,,&\hskip .2cm  x^i & = & \frac{x'^i}{1-ct'}\,,\\
\phi & = & (1-ct')^{d/2}\phi'\,,&\hskip .2cm M & = & M'+\frac{c}{2}\frac{x'^ix'^i}{1-ct'}\,.
\end{array}
\end{equation}
Under the remaining symmetries $H$, $P_i$, $G_i$, $N$ and $J_{ij}$ the field $\phi$ transforms as a scalar and $M$ transforms as described before, i.e. as in \eqref{eq:KE} with $z=2$.

\noindent\textbf{Schr\"odinger Invariance from Lifshitz Isometries.} It is a well known fact that the free Schr\"odinger equation obtained by varying \eqref{eq:Schaction2} with respect to $\psi$ for $V=0$ is left invariant under the Schr\"odinger group. The way in which the Schr\"odinger symmetries are realized on $\psi$ is via a projective UIR of the Schr\"odinger group without $N$ \cite{Niederer:1972zz}. However the way in which they are realized on $\phi$ is quite different. As detailed earlier we have to consider two solutions of \eqref{eq:noNP}, equations \eqref{eq:choice1M} and \eqref{eq:choice2M}, and for each of these choices $\phi$ transforms as a UIR representation of the Lifshitz subalgebras spanned by $H$, $P_i$, $J_{ij}$, $D$ and by $K$, $G_i$, $J_{ij}$, $D$, respectively. If we take $M=\text{cst}$ then $\phi$ transforms projectively under the $K$, $G_i$ transformations and vice versa when we take $M=\tfrac{x^ix^i}{2t}$ the field $\phi$ transforms projectively under the $H$ and $P_i$ transformations.

Instead of working with one projective UIR $\psi$ of the Schr\"odinger group without $N$ we use two UIRs $\phi$ of two Lifshitz subgroups (one for each $M$). These Lifshitz subgroups are related to each other by the outer automorphisms of the $z=2$ Schr\"odinger algebra
\begin{equation}
\left(H, P_i, D, J_{ij}\right) \leftrightarrow \left(-K, G_i, -D, J_{ij}\right)\,.
\end{equation}
Hence the space-time symmetries are always given by a Lifshitz algebra of Killing vectors. The $N$ generator is not realized as a space-time symmetry. When we include $N$ it will be odd under the $\mathbb{Z}_2$ outer automorphism.

The on-shell conserved currents related to $H$, $P_i$, $J_{ij}$, $D$ invariance of \eqref{eq:SchactionflatNC} for $M=\text{cst}$ can be written, using \eqref{eq:EMtensor+masscurrent}--\eqref{eq:virialcurrent} evaluated for a flat NC space-time, as
\begin{equation}
\begin{array}{rclrcl}
\partial_\mu\left(H^\nu T^\mu{}_\nu\right) & = & 0\,,&\hskip .2cm \partial_\mu\left(P_i^\nu T^\mu{}_\nu\right) & = & 0\,,\\
\partial_\mu\left(J_{ij}^\nu T^\mu{}_\nu\right) & = & 0\,,&\hskip .2cm \partial_\mu\left(D^\nu T^\mu{}_\nu-V^\mu\right) & = & 0\,,\\
\end{array}
\end{equation}
where the Killing vectors are given in \eqref{eq:HPJD}. When $M$ is given by \eqref{eq:choice2M} the on-shell conserved currents are
\begin{equation}
\begin{array}{rclrcl}
\partial_\mu\left(K^\nu T^\mu{}_\nu-tV^\mu\right) & = & 0\,,&\hskip .2cm \partial_\mu\left(G_i^\nu T^\mu{}_\nu\right) & = & 0\,,\\
\partial_\mu\left(J_{ij}^\nu T^\mu{}_\nu\right) & = & 0\,,&\hskip .2cm \partial_\mu\left(D^\nu T^\mu{}_\nu-V^\mu\right) & = & 0\,,
\end{array}
\end{equation}
where the (conformal) Killing vectors are given in \eqref{eq:KGJD}.

We have concentrated our attention to space-time symmetries. The $N$ generator acts on field space and when we include such transformations the algebra becomes the full Schr\"odinger algebra including the central element $N$ and the commutator $[P_i,G_j]=\delta_{ij}N$. To realize this algebra on the field $\phi$ we need to add additional terms to the $K$ and $G_i$ generators of \eqref{eq:KGJD}, that act only on field space (see for example \cite{Nishida:2007pj}).

We would like to stress that in general when writing down a field theory on a flat NC background there may also be cases where we cannot remove $M$ from the action, e.g. when the potential in \eqref{eq:SchactionflatNC} breaks the $U(1)$ symmetry of the model or when we consider an action for a real scalar coupled to a TNC background such as the Lifshitz model
\begin{equation}\label{eq:Lifshitzmodel}
S = \int d^{d+1}xe\left(\frac{1}{2}\left(\hat v^\mu\partial_\mu\phi\right)^2-\frac{\kappa}{2}\left(h^{\mu\nu}\nabla_\mu\partial_\nu\phi\right)^2\right)\,.
\end{equation}
If we specify this model to the case of a flat NC space-time the action will depend on what we take for $M$ and hence we can at most (in the sense of the largest number of symmetries) obtain the Lifshitz algebra. We will now proceed to study field theories defined on a TNC background that are defined holographically.

\noindent\textbf{Holography with Lifshitz Bulk Geometry.} We have seen that the prototype of Schr\"odinger invariant field theory, namely the action \eqref{eq:Schaction2} leading to the Schr\"odinger equation, is based on flat NC geometry with Lifshitz conformal Killing vectors. In order to study holography for Schr\"odinger invariant systems a natural starting point is thus to take a Lifshitz bulk space-time geometry. In fact in \cite{Hartong:2014oma} we have shown that asymptotically locally Lifshitz space-times provide a set of sources that describe TNC geometry, that transform under the Schr\"odinger algebra in the sense of \cite{Bergshoeff:2014uea}, i.e. making local translations equivalent to diffeomorphisms, and that these sources couple to vevs whose Ward identities are organized by the Schr\"odinger algebra. Here we re-examine this claim by studying the symmetries of the Lifshitz vacuum.

Consider exact Lifshitz space-times in coordinates such that it admits a flat NC boundary as defined in \eqref{eq:flatNC}. These bulk geometries thus depend on $M$ only. We will construct Lifshitz bulk geometries that correspond to the choices \eqref{eq:choice1M} and \eqref{eq:choice2M} using symmetry arguments. 

When $M=\text{cst}$ the boundary conformal Killing vectors are \eqref{eq:HPJD}. Of these only $D$ is an actual conformal Killing vector. The other three are Killing vectors. This suggests that also in the bulk $H$, $P_i$ and $J_{ij}$ will be Killing vectors without modification whereas for $D$ we take the bulk Killing vector to include a radial component so that now
\begin{equation}\label{eq:bulkD}
D=zt\partial_t+x\partial_x+y\partial_y+r\partial_r\,.
\end{equation}
This is of course the well-known dilatation generator of the 4D bulk Lifshitz space-time. Hence the most general metric respecting these symmetries is  the familiar Lifshitz metric
\begin{equation}
ds^2=-\frac{dt^2}{r^{2z}}+2C\frac{drdt}{r^{z+1}}+\frac{dr^2}{r^2}+\frac{1}{r^2}\left(dx^2+dy^2\right)\,,
\end{equation}
where the constant $C$ can be removed by choosing a new coordinate $\bar t$ given by $\bar t=t-\frac{C}{z}r^z$. We thus conclude that the standard form of the Lifshitz space-time metric corresponds to a flat NC boundary with $M=\text{cst}$.

Let us now construct the bulk dual to a flat NC boundary with $M=(x^2+y^2)/2t$. For this purpose we need to extend the boundary conformal Killing vectors $D$ and $K$ into the bulk where they become bulk Killing vectors while leaving the boundary Killing vectors $G_i$ and $J_{ij}$ unaltered. We are going to use the same set of coordinates as before so we again take for the bulk realization of $D$ the expression \eqref{eq:bulkD}. In order to obey the commutation relations \eqref{eq:Lifshitzalgebra2} we need to take for $K$ the bulk expression
\begin{equation}\label{eq:bulkK}
K=t^z\partial_t+t^{z-1}\left(x\partial_x+y\partial_y+r\partial_r\right)\,.
\end{equation}
We thus demand that there exists a 4D bulk metric that has the Killing vectors \eqref{eq:bulkD}, \eqref{eq:bulkK}, as well as $G_i$ and $J_{ij}$ given in \eqref{eq:KGJD}. The resulting bulk metric reads
\begin{eqnarray}
ds^2 & = & \left(-\frac{1}{r^{2z}}-\frac{2C}{r^{z}t}+\frac{1}{t^{2}}\right)dt^2+2\left(-\frac{1}{rt}+\frac{C}{r^{z+1}}\right)drdt\nonumber\\
&&+\frac{dr^2}{r^2}+\frac{1}{r^2}\left[\left(dx-\frac{x}{t}dt\right)^2+\left(dy-\frac{y}{t}dt\right)^2\right]\,,\label{eq:Lifm0-choice2}
\end{eqnarray}
where $C$ is a constant. This is thus the vacuum bulk geometry dual to a flat NC boundary with $M=(x^2+y^2)/2t$. Note that the $t,x,y$ dependence of the metric can be written in terms of $M$ and its derivatives via
$\partial_x M=x/t$, $\partial_y M=y/t$ and $(\partial_x^2+\partial_y^2)M=2/t$. Another noteworthy feature of this metric is that even for $C=0$ it is not in radial gauge due to the $drdt$ term. 
The algebra of Killing vectors that this metric possesses is isomorphic to the Lifshitz algebra and hence we should be able to show that this metric is diffeomorphic to the Lifshitz space-time. Consider the following coordinate transformation
\begin{equation}\label{eq:coordinatetrafo}
\bar t = \frac{1}{1-z}t^{1-z}\,,\hskip .3cm \bar x = \frac{x}{t}\,,\hskip .3cm \bar y = \frac{y}{t}\,,\hskip .3cm \bar r = \frac{r}{t}\,.
\end{equation}
This takes the metric \eqref{eq:Lifm0-choice2} to the following form
\begin{equation}
ds^2 = -\frac{d\bar t^2}{\bar r^{2z}}+2C\frac{d\bar rd\bar t}{\bar r^{z+1}}+\frac{d\bar r^2}{\bar r^2}+\frac{1}{\bar r^2}\left(d\bar x^2+d\bar y^2\right)\,.
\end{equation}
The Killing vectors in this coordinate system read
\begin{eqnarray}
K & = & \partial_{\bar t}\,,\hskip .2cm G_{\bar x} = \partial_{\bar x}\,,\hskip .2cm G_{\bar y} = \partial_{\bar y}\,,\hskip .2cm J_{\bar x\bar y} = \bar x\partial_{\bar y}-\bar y\partial_{\bar x}\,,\nonumber\\
D & = & -(z-1)\left[z\bar t\partial_{\bar t}+\bar x\partial_{\bar x}+\bar y\partial_{\bar y}+\bar r\partial_{\bar r}\right]\,.
\end{eqnarray}
Hence switching, on the boundary, from $M=\text{cst}$ to $M=(x^2+y^2)/2t$ corresponds in the bulk to a diffeomorphism. In \cite{Hartong:2015wxa} we show that this diffeomorphism can be turned into a Penrose--Brown--Henneaux transformation by adding terms that are subleading in $r$ to \eqref{eq:coordinatetrafo} so that a bulk Lifshitz space-time contains two versions of the Lifshitz algebra corresponding to a flat NC boundary with different realizations of $M$ that are related by a local symmetry of the holographic model. 

A bulk Lifshitz space-time has the same properties as flat NC space-time. Since it is possible to construct theories on flat NC space-time that have more symmetries than there are conformal Killing vectors it should be possible to choose matter fields on a Lifshitz space-time that possess global Schr\"odinger symmetries. In \cite{Hartong:2015wxa} we show that this is indeed the case.


\noindent\textbf{Discussion.} This work together with \cite{Hartong:2014oma,Bergshoeff:2014uea} shows that Lifshitz holography is not an implementation of a holographic duality between a gravitational theory on asymptotically locally Lifshitz space-times and Lifshitz invariant field theories. The dual field theory can have more, in particular global Schr\"odinger invariance. In fact in \cite{Christensen:2013lma,Christensen:2013rfa} a holographic Lifshitz model has been obtained by a reduction from an asymptotically AdS space-time on a circle that becomes null on the AdS boundary. We are thus reducing a CFT on a null circle which is expected to lead to Schr\"odinger and not just Lifshitz symmetries. 

This is a serious shift in perspective that is expected to have important consequences for applying holography to strongly coupled systems with non-relativistic symmetries 
\cite{Son:2008ye,Balasubramanian:2008dm,Kachru:2008yh,Taylor:2008tg}. In some sense we must rethink what we use Lifshitz space-times for. It would be interesting to extend this discussion to black holes and apply the ideas of fluid/gravity to study fluids on a TNC background. Further we expect these results to point the way towards a holographic description of the effective field theory for the quantum Hall effect.

{\bf Note added:}
While this letter was being finalized, the preprint \cite{Jensen:2014aia} appeared on the arXiv, which appears to have some overlap with our results regarding coupling to TNC backgrounds. 

\noindent\textbf{Acknowledgements.} We would like to thank Eric Bergshoeff, Geoffrey Comp\`ere, Simon Ross, Jan Rosseel and especially Jay Armas, Matthias Blau, Jan de Boer and Kristan Jensen for many valuable discussions. The work of JH is supported in part by the Danish National Research Foundation project ``Black holes and their role in quantum gravity''. The work of EK was supported in part by European Union's Seventh Framework Programme under grant agreements (FP7-REGPOT-2012-2013-1) no 316165,
PIF-GA-2011-300984, the EU program ``Thales'' MIS 375734, by the European Commission under the ERC Advanced Grant BSMOXFORD 228169 and was also co-financed by the European Union (European Social Fund, ESF) and Greek national funds through the Operational Program ``Education and Lifelong Learning'' of the National Strategic Reference Framework (NSRF) under ``Funding of proposals that have received a positive evaluation in the 3rd and 4th Call of ERC Grant Schemes''. The work of NO is supported in part by  Danish National Research Foundation project ``New horizons in particle and condensed matter physics from black holes".  JH wishes to thank CERN for its hospitality and financial support.

\addcontentsline{toc}{section}{References}

\vspace{-.2cm}

\bibliography{SchandLif-letter}

\end{document}